*Structural bioinformatics*

# COVID-19 Docking Server: A meta server for docking small molecules, peptides and antibodies against potential targets of COVID-19

Ren Kong[1,*], Guangbo Yang[1], Rui Xue[1], Ming Liu[2], Feng Wang[3], Jianping Hu[4], Xiaoqiang Guo[4], Shan Chang[1,*]

[1]Institute of Bioinformatics and Medical Engineering, School of Electrical and Information Engineering, Jiangsu University of Technology, Changzhou 213001, China

[2]Beijing New BioConcepts Biotech Co., Ltd., Beijing 101111, China

[3]School of Information Science & Engineering, Changzhou University, Changzhou 213164, China

[4]Key Laboratory of Medicinal and Edible Plants Resources Development of Sichuan Education Department, College of Pharmacy and Biological Engineering, Sichuan Industrial Institute of Antibiotics, Chengdu University, Chengdu 610106, China

*To whom correspondence should be addressed.

**Abstract**
**Motivation:** The coronavirus disease 2019 (COVID-19) caused by a new type of coronavirus has been emerging from China and led to thousands of death globally since December 2019. Despite many groups have engaged in studying the newly emerged virus and searching for the treatment of COVID-19, the understanding of the COVID-19 target-ligand interactions represents a key challenge. Herein, we introduce COVID-19 Docking Server, a web server that predicts the binding modes between COVID-19 targets and the ligands including small molecules, peptides and antibodies.
**Results:** Structures of proteins involved in the virus life cycle were collected or constructed based on the homologs of coronavirus, and prepared ready for docking. The meta platform provides a free and interactive tool for the prediction of COVID-19 target-ligand interactions and following drug discovery for COVID-19.
**Availability:** http://ncov.schanglab.org.cn
**Contact:** rkong@jsut.edu.cn or schang@jsut.edu.cn
**Supplementary information:** Supplementary data are available at *Bioinformatics* online.

## 1 Introduction

According the situation report from World Health Organization (WHO), 7,039,918 cases of coronavirus disease 2019 (COVID-19) and 404,396 deaths have been reported all around the world till June 9th, 2020. The first COVID-19 case was reported in Wuhan city, Hubei province of China in December 2019. The pathogen caused the disease was soon identified as a novel coronavirus, which belongs to the genus Betacoronavirus and is closely related to severe acute respiratory syndrome coronavirus (SARS-CoV) with 89.1% nucleotide similarity in the viral genome (Jiang, et al., 2020; Wu, et al., 2020). Later on, it was named as severe acute respiratory syndrome coronavirus 2 (SARS-CoV-2) by the International Committee on the Taxonomy of Viruses. Tremendous efforts have been done to study the newly emerged virus and find potent drugs for clinical usage. Wang *et. al* from Wuhan Institute of Virology screened some of the FDA approved anti-virus or anti-infection drugs and found that remdesivir and chloroquine could effectively inhibit the virus in cell based assay with $EC_{50}$ of 0.77 and 1.13 μM, respectively (Wang, et al., 2020). Several clinical trials are ongoing for the treatment of COVID-19. However, no drug or vaccine has yet been approved.

In a very short time, the structures of functional proteins essential for SARS-CoV-2 were achieved by groups in China, including the main protease and spike protein in binding with angiotensin-converting enzyme 2 (ACE2). Until May 20, totally 181 structures were released in the Protein Data Bank, which related to 11 different kinds of SARS-CoV-2 proteins. Although the structures of other proteins remain unknown, the high amino acid identity between SARS-CoV-2 and SARS-CoV (77.2%) enables that it is possible to establish the homology modeled structures of SARS-CoV-2 based on the protein structures of SARS-CoV with high confidence. I-TASSER group provided 24 high quality predicted structures of SARS-CoV-2 to download (Zhang, et al., 2020). Based on the these structure information, a web-server, COVID-19 Docking Server, was constructed to facilitate people to evaluate the binding modes and binding affinities between the targets and small molecules, peptides as well as antibodies. By launching the server, we hope to provide a free and easy to use tool for people who is interested in drug discovery against COVID-19.

## 2 Implementation

The COVID-19 Docking Server web interface is written in PHP and HTML, and JSMol (http://jmol.sourceforge.net/) is used for molecular visualization on the results pages. Totally, 27 targets essential in virus life cycle are prepared and available for docking on the website (Supplementary Information, Fig. S1 and S2), including spike protein, membrane protein, envelop small membrane protein, nucleocapsid protein, main protease, papain-like protease, nsp3 (207-379), RNA dependent RNA polymerase (RdRp, nsp12/7/8 complex) , nsp7, nsp8, nsp12, helicase, nsp14, nsp15 (endoribonuclease), nsp10, nsp16, nsp16/10 (2'-O-methyltransferase), nsp1, nsp2, nsp4, nsp6, nsp9, ORF3A, ORF6, ORF7A, ORF8, ORF10. The flow chart of the COVID-19 Docking Server is shown in Fig. S3.

For small molecule docking, Autodock Vina (Trott and Olson, 2010) is used as docking engine. Open Babel was used for format transformation or 3D coordinate generation for the uploaded files (O'Boyle, et al., 2011). For experimental complex structures, the docking box was defined on the center of native ligand with 30Å×30Å×30Å in length to include the residues of entire cavity. For homology modeled structures, it is defined according to the information of active sites or binding sites of its homologs of SARS-CoV (detailed in Supplementary Information). All the parameters are set as default and we provide exhaustiveness value options for the user to choose the docking precision. Except for Autodock Vina score, we also provide an alternative scoring function as complement to evaluate the binding affinities for small molecules (Li, et al., 2016).

For peptide and antibody docking, CoDockPP (Kong, et al., 2019) is used as docking engine. CoDockPP program provides a multistage FFT-based strategy for both global docking and site-specific docking. An angle interval of 15º is used for rotational sampling, and a spacing of 1.2 Å is adopted for fast Fourier transform (FFT) translational search. Finally, the top binding modes are clustered with ligand root mean square deviations (L_RMSD) cutoff of 3.0 Å in global docking and 2.0 Å in site-specific docking.

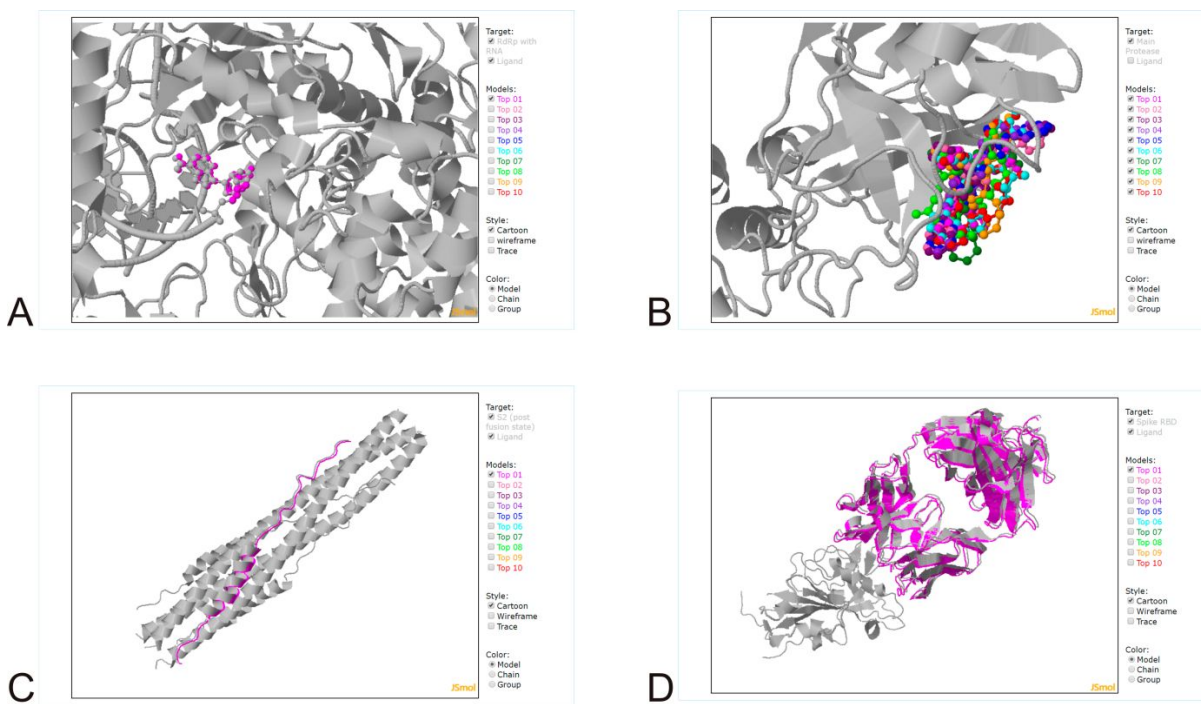

**Fig. 1. COVID-19 Docking Server outputs.** (A) An example of small molecule docking. The RdRp docked with its inhibitor. The experimental structure of RdRp with its inhibitor is colored in gray. The predicted structure of Top 1 is colored in pink. (B) An example of small molecule batch docking. The crystal structure of main protease is colored gray and the predicted binding poses of 11 inhibitors are shown in different colors. (C) An example of peptide docking. The post-fusion state of S2 segment of spike protein docked with a helical

peptide. The 6-helices structure of S2 is colored in gray. The predicted structure of Top 1 is colored in pink. (D) An example of antibody docking. The spike protein docked with antibody. The spike protein is colored in gray and the predicted structure of Top 1 is colored in pink.

## 3  Usage

### 3.1  Usage of small molecule docking

For each job of small molecule docking, the user can choose one of the COVID-19 targets. The small molecules should be uploaded in strict smi, mol2 or sdf formats. Two computational types are provided with the default computational type as "Docking" (D) to dock single small molecule to the target, and the other option "Batch Docking" (B) to dock a number of small molecules to the target. The top 10 models are visualized in 3D by JSmol and the user can view and download the docking results from the result page.  Here, we take RdRp and its inhibitor triphosphate form of remdesivir as an example for small molecule docking (Yin, et al., 2020). The "Docking" mode was selected and the mol2 format of its inhibitor in PDB 7BV2 was uploaded as ligand. The predicted binding energy of Top 1 ranked binding mode was -8.3 kcal/mol, and the L_RMSD was 1.97 Å. As shown in Fig. 1A, the best binding mode of the inhibitor is very similar to the native ligand in the experimental structure. In addition, if users have a set of small molecules and want to explore the changes in predicted binding affinity with various chemical substitutions, they can use batch docking mode as shown in Fig. 1B.

### 3.2  Usage of peptide or antibody docking

Similar to small molecule docking, the user needs to choose the COVID-19 protein target and uploads ligand protein (peptide or antibody) in strict pdb format for peptide or antibody docking. The default mode is globe docking and the user also could input residues for site specific docking. If the user defined constraint residues for the receptor and ligand simultaneously, then he needs to choose the constraint type: ambiguous constraints    or multiple constraints.    When the "ambiguous constraints" is selected, the conformations are retained with at least one selected residue on the interface. When the "multiple constraints" is selected, the conformations are retained with both of the residues on the interface. We use the post-fusion state of S2 segment of spike protein as an example for peptide docking, and RBD of spike protein as an example for antibody docking. The 5-helices structure from 6LXT was used as receptor and the other helix was used as ligand peptide (Xia, et al., 2020). By using the default global searching, the L_RMSD for Top 1 binding mode was 0.93 Å (Fig. 1C). The antibody structure from 7BZ5 (Wu, et al., 2020) were extracted and docked to RBD by using global searching option. And the L_RMSD for Top 1 binding mode was 1.14 Å (Fig. 1D).

To evaluation the docking protocols, the re-dock experiments were conducted for those targets with experimental complex structure available. Most of the complex structures were reproduced by using the docking procedure on the web server (Table S1 & S2 in Supplementary Information). Since online on March 5, 2020, the COVID-19 server has completed > 7,000 jobs from about 400 unique users in three months. As the continuously increasing infections of SARS-CoV-2 in a number of countries all around the world, more researchers and groups are involved in the study of this virus and the demand of prediction for target-ligand complex structures may increase accordingly.

## 4  Conclusions

An online meta server, COVID-19 Docking Server, was constructed to predict the binding modes between the targets of COVID-19 and its potential ligands by implement of Autodock Vina and CoDockPP as docking engines. The server provides a user-friendly interface and binding mode visualization for the results, which makes it a useful tool for drug discovery of COVID-19.

**Acknowledgements**

We would like to thank Zihe Rao's group in Shanghai University of Technology and Xinquan Wang's group in Tsinghua University for their generously sharing the structures of main protease, spike protein and ACE2 protein. We also want to thank www.yaozh.com to provide the financial support to rent the server from Aliyun.

**Funding**

This work has been supported by the National Natural Science Foundation of China (81603152), Six Talent Peaks Project in Jiangsu Province (Grant No. 2016-XYDXXJS-020) and the Key Project of Science and Technology Department of Sichuan Province (2020YFS0006).

*Conflict of Interest:* none declared.